\pdfoutput=1
\documentclass[12pt,a4paper]{article}
\usepackage{ifthen} % for conditional statements
\newboolean{pdflatex}
\setboolean{pdflatex}{true} % False for eps figures 

\newboolean{articletitles}
\setboolean{articletitles}{true} % False removes titles in references

\newboolean{uprightparticles}
\setboolean{uprightparticles}{false} %True for upright particle symbols

\def\paperauthors{LHCb collaboration} % Leave as is for PAPER, CONF and FIGURE
\def\paperasciititle{First observation of the Cabibbo-suppressed decay Bc+ to JPsi D*+} % Set ASCII title here !! MAKE sure it's only ASCII characters !! 
\def\papertitle{First observation of the Cabibbo-suppressed decay \BcJpsiDst} % Latex formatted title
\def\paperkeywords{{High Energy Physics}, {LHCb}} % Comma separated list
\def\papercopyright{\the\year\ CERN for the benefit of the LHCb collaboration} % new since 9/Apr/2018
\def\paperlicence{CC BY 4.0 licence}
\def\paperlicenceurl{https://creativecommons.org/licenses/by/4.0/}

\newif\ifEnableSectionTOCLinks
\EnableSectionTOCLinksfalse % deactivated

\usepackage{caption}
\usepackage{subcaption}
\usepackage{booktabs}

\usepackage[top=1in, bottom=1.25in, left=1in, right=1in]{geometry}

\usepackage{microtype}
\usepackage{lineno}  % for line numbering during review
\usepackage{xspace} % To avoid problems with missing or double spaces after
\usepackage{caption} %these three command get the figure and table captions automatically small

\usepackage{graphicx}  % to include figures (can also use other packages)
\usepackage{color}
\usepackage{colortbl}
\graphicspath{{./figs/}} % Make Latex search fig subdir for figures
\usepackage{amsmath} % Adds a large collection of math symbols
\usepackage{amssymb}
\usepackage{amsfonts}
\usepackage{upgreek} % Adds in support for greek letters in roman typeset

\newcommand*\patchAmsMathEnvironmentForLineno[1]{%
\expandafter\let\csname old#1\expandafter\endcsname\csname #1\endcsname
\expandafter\let\csname oldend#1\expandafter\endcsname\csname
end#1\endcsname
 \renewenvironment{#1}%
   {\linenomath\csname old#1\endcsname}%
   {\csname oldend#1\endcsname\endlinenomath}%
}
\newcommand*\patchBothAmsMathEnvironmentsForLineno[1]{%
  \patchAmsMathEnvironmentForLineno{#1}%
  \patchAmsMathEnvironmentForLineno{#1*}%
}
\AtBeginDocument{%
\patchBothAmsMathEnvironmentsForLineno{equation}%
\patchBothAmsMathEnvironmentsForLineno{align}%
\patchBothAmsMathEnvironmentsForLineno{flalign}%
\patchBothAmsMathEnvironmentsForLineno{alignat}%
\patchBothAmsMathEnvironmentsForLineno{gather}%
\patchBothAmsMathEnvironmentsForLineno{multline}%
\patchBothAmsMathEnvironmentsForLineno{eqnarray}%
}

\usepackage[pdftex,
            pdfauthor={\paperauthors},
            pdftitle={\paperasciititle},
            pdfkeywords={\paperkeywords}]{hyperref}
\usepackage{hyperxmp}
\hypersetup{
    pdfcopyright={Copyright (C) \papercopyright},
    pdflicenseurl={\paperlicenceurl}
}
\usepackage[colorinlistoftodos,textsize=scriptsize]{todonotes}

\usepackage[bottom,flushmargin,hang,multiple]{footmisc}

\usepackage[all]{hypcap} % Internal hyperlinks to floats.

\usepackage{xspace} 
\usepackage{upgreek}

\def\lhcb   {\mbox{LHCb}\xspace}
\def\atlas  {\mbox{ATLAS}\xspace}

\def\MagUp {\mbox{\em Mag\kern -0.05em Up}\xspace}

\ifthenelse{\boolean{uprightparticles}}%
{

 \def\Pmu         {\ensuremath{\upmu}\xspace}

 \def\Ppi         {\ensuremath{\uppi}\xspace}

 \def\Ppsi        {\ensuremath{\uppsi}\xspace}

 \def\PDelta      {\ensuremath{\Delta}\xspace}                 
 \def\PXi         {\ensuremath{\Xi}\xspace}                 
 \def\PLambda     {\ensuremath{\Lambda}\xspace}                 
 \def\PSigma      {\ensuremath{\Sigma}\xspace}                 
 \def\POmega      {\ensuremath{\Omega}\xspace}                 
 \def\PUpsilon    {\ensuremath{\Upsilon}\xspace}
 \let\oldPi\Pi
 \def\PPi         {\ensuremath{\oldPi}\xspace}

 \def\PB      {\ensuremath{\mathrm{B}}\xspace}                 
 \def\PD      {\ensuremath{\mathrm{D}}\xspace}                 
 \def\PJ      {\ensuremath{\mathrm{J}}\xspace}                 
 \def\PK      {\ensuremath{\mathrm{K}}\xspace}                 
 \def\Pb      {\ensuremath{\mathrm{b}}\xspace}                 
 \def\Pc      {\ensuremath{\mathrm{c}}\xspace}

 \def\Ps      {\ensuremath{\mathrm{s}}\xspace}

 \def\thebaroffset{0.0em}
}
{

 \def\Pmu         {\ensuremath{\mu}\xspace}

 \def\Ppi         {\ensuremath{\pi}\xspace}

 \def\Ppsi        {\ensuremath{\psi}\xspace}                 
                  
 \mathchardef\PDelta="7101
 \mathchardef\PXi="7104
 \mathchardef\PLambda="7103
 \mathchardef\PSigma="7106
 \mathchardef\POmega="710A
 \mathchardef\PUpsilon="7107
 \mathchardef\PPi="7105
 \def\PB      {\ensuremath{B}\xspace}                 
 \def\PD      {\ensuremath{D}\xspace}                 
 \def\PJ      {\ensuremath{J}\xspace}                 
 \def\PK      {\ensuremath{K}\xspace}                 
 \def\Pb      {\ensuremath{b}\xspace}                 
 \def\Pc      {\ensuremath{c}\xspace}

 \def\Ps      {\ensuremath{s}\xspace}

 \def\thebaroffset{0.18em}
}
\newcommand{\offsetoverline}[2][\thebaroffset]{\kern #1\overline{\kern -#1 #2}}%

\makeatletter
\ifcase \@ptsize \relax% 10pt
  \newcommand{\miniscule}{\@setfontsize\miniscule{4}{5}}% \tiny: 5/6
\or% 11pt
  \newcommand{\miniscule}{\@setfontsize\miniscule{5}{6}}% \tiny: 6/7
\or% 12pt
  \newcommand{\miniscule}{\@setfontsize\miniscule{5}{6}}% \tiny: 6/7
\fi
\makeatother

\DeclareRobustCommand{\optbar}[1]{\shortstack{{\miniscule (\rule[.5ex]{1.25em}{.18mm})}
  \\ [-.7ex] $#1$}}

\def\mumu       {{\ensuremath{\Pmu^+\Pmu^-}}\xspace}

\def\squark    {{\ensuremath{\Ps}}\xspace}

\def\cquark    {{\ensuremath{\Pc}}\xspace}

\def\bquark    {{\ensuremath{\Pb}}\xspace}
\def\bquarkbar {{\ensuremath{\overline \bquark}}\xspace}
\def\bbbar     {{\ensuremath{\bquark\bquarkbar}}\xspace}

\def\pion   {{\ensuremath{\Ppi}}\xspace}

\def\pip    {{\ensuremath{\pion^+}}\xspace}
\def\pim    {{\ensuremath{\pion^-}}\xspace}

\def\kaon    {{\ensuremath{\PK}}\xspace}
\def\KorKbar {\kern \thebaroffset\optbar{\kern -\thebaroffset \PK}{}\xspace}

\def\Kp      {{\ensuremath{\kaon^+}}\xspace}
\def\Km      {{\ensuremath{\kaon^-}}\xspace}

\def\D       {{\ensuremath{\PD}}\xspace}

\def\DorDbar {\kern \thebaroffset\optbar{\kern -\thebaroffset \PD}\xspace}
\def\Dz      {{\ensuremath{\D^0}}\xspace}

\def\Dp      {{\ensuremath{\D^+}}\xspace}
\def\Dm      {{\ensuremath{\D^-}}\xspace}

\def\DpDm    {\ensuremath{\Dp {\kern -0.16em \Dm}}\xspace}

\def\Dstarp  {{\ensuremath{\D^{*+}}}\xspace}

\def\Ds      {{\ensuremath{\D^+_\squark}}\xspace}
\def\Dsp     {{\ensuremath{\D^+_\squark}}\xspace}

\def\Dssp    {{\ensuremath{\D^{*+}_\squark}}\xspace}

\def\B       {{\ensuremath{\PB}}\xspace}

\def\BorBbar {\kern \thebaroffset\optbar{\kern -\thebaroffset \PB}\xspace}

\def\Bd      {{\ensuremath{\B^0}}\xspace}

\def\BdorBdbar {\kern \thebaroffset\optbar{\kern -\thebaroffset \Bd}\xspace}
\def\Bu      {{\ensuremath{\B^+}}\xspace}

\def\Bp      {{\ensuremath{\Bu}}\xspace}

\def\Bs      {{\ensuremath{\B^0_\squark}}\xspace}

\def\BsorBsbar {\kern \thebaroffset\optbar{\kern -\thebaroffset \Bs}\xspace}
\def\Bc      {{\ensuremath{\B_\cquark^+}}\xspace}

\def\jpsi     {{\ensuremath{{\PJ\mskip -3mu/\mskip -2mu\Ppsi}}}\xspace}

\def\Y#1S{\ensuremath{\PUpsilon{(#1S)}}\xspace}

\def\Lz          {{\ensuremath{\PLambda}}\xspace}

\def\LorLbar     {\kern \thebaroffset\optbar{\kern -\thebaroffset \PLambda}\xspace}

\def\Lc          {{\ensuremath{\Lz^+_\cquark}}\xspace}

\newcommand{\decay}[2]{\ensuremath{\mathinner{#1\!\to #2}}\xspace}

\def\to                 {\ensuremath{\rightarrow}\xspace}

\def\AT#1     {\ensuremath{A_{\mathrm{T}}^{#1}}\xspace}           % 2

\def\C#1      {\ensuremath{\mathcal{C}_{#1}}\xspace}                       % 9
\def\Cp#1     {\ensuremath{\mathcal{C}_{#1}^{'}}\xspace}                    % 7
\def\Ceff#1   {\ensuremath{\mathcal{C}_{#1}^{\mathrm{(eff)}}}\xspace}        % 9  
\def\Cpeff#1  {\ensuremath{\mathcal{C}_{#1}^{'\mathrm{(eff)}}}\xspace}       % 7
\def\Ope#1    {\ensuremath{\mathcal{O}_{#1}}\xspace}                       % 2
\def\Opep#1   {\ensuremath{\mathcal{O}_{#1}^{'}}\xspace}                    % 7

\newcommand{\nospaceunit}[1]{\ensuremath{\text{#1}}}       
\newcommand{\aunit}[1]{\ensuremath{\text{\,#1}}}       
\newcommand{\tev}{\aunit{Te\kern -0.1em V}\xspace}
\newcommand{\gev}{\aunit{Ge\kern -0.1em V}\xspace}
\newcommand{\mev}{\aunit{Me\kern -0.1em V}\xspace}
\newcommand{\kev}{\aunit{ke\kern -0.1em V}\xspace}
\newcommand{\ev}{\aunit{e\kern -0.1em V}\xspace}
 
\newcommand{\mevc}{\ensuremath{\aunit{Me\kern -0.1em V\!/}c}\xspace}
\newcommand{\gevc}{\ensuremath{\aunit{Ge\kern -0.1em V\!/}c}\xspace}
\newcommand{\mevcc}{\ensuremath{\aunit{Me\kern -0.1em V\!/}c^2}\xspace}
\newcommand{\gevcc}{\ensuremath{\aunit{Ge\kern -0.1em V\!/}c^2}\xspace}
\def\mum  {\ensuremath{\,\upmu\nospaceunit{m}}\xspace}

\def\fb   {\ensuremath{\aunit{fb}}\xspace}
\def\invfb   {\ensuremath{\fb^{-1}}\xspace}

\def\gsim{{~\raise.15em\hbox{$>$}\kern-.85em
          \lower.35em\hbox{$\sim$}~}\xspace}
\def\lsim{{~\raise.15em\hbox{$<$}\kern-.85em
          \lower.35em\hbox{$\sim$}~}\xspace}

\def\pt         {\ensuremath{p_{\mathrm{T}}}\xspace}

\def\ptot       {\ensuremath{p}\xspace}

\def\bcvegpy    {\mbox{\textsc{Bcvegpy}}\xspace}

\def\evtgen     {\mbox{\textsc{EvtGen}}\xspace}

\def\geant      {\mbox{\textsc{Geant4}}\xspace}

\def\photos     {\mbox{\textsc{Photos}}\xspace}

\def\pythia     {\mbox{\textsc{Pythia}}\xspace}

\def\tell1  {TELL1\xspace}
\def\ukl1   {UKL1\xspace}

\newcommand{\lhcborcid}[1]{\href{https://orcid.org/#1}{\hspace*{0.1em}\raisebox{-0.45ex}{\includegraphics[width=1em]{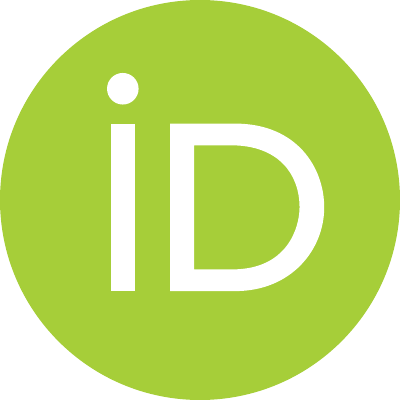}}}}

\hypersetup{
  colorlinks   = true, %Colours links instead of ugly boxes
  urlcolor     = blue, %Colour for external hyperlinks
  linkcolor    = blue, %Colour of internal links
  citecolor    = red   %Colour of citations
}

\ifEnableSectionTOCLinks
    \usepackage[explicit]{titlesec} % to change headings
    
    \let\oldcontentsline\contentsline
    \renewcommand

    \titleformat{\section}{\normalfont\Large\bf}{\hyperlink{tocsection.\thesection}{{\thesection} \parbox[t]{\dimexpr\textwidth-1pc}{#1}}}{1pc}{}

    \titleformat{\subsection}{\normalfont\bf}{\hyperlink{tocsubsection.\thesubsection}{{\thesubsection} \parbox[t]{\dimexpr\textwidth-1pc}{#1}}}{1pc}{}

    \titleformat{name=\section,numberless}[display]{}{}{0pt}{\normalfont\Huge\bfseries #1}
\fi

\usepackage{cite} % Allows for ranges in citations
\usepackage{LHCb/mciteplus}
\usepackage{xspace} 
\usepackage{upgreek}

\def\lhcb   {\mbox{LHCb}\xspace}

\def\Pmu         {\ensuremath{\mu}\xspace}

\def\mumu       {{\ensuremath{\Pmu^+\Pmu^-}}\xspace}

\def\pip    {{\ensuremath{\pion^+}}\xspace}
\def\pim    {{\ensuremath{\pion^-}}\xspace}

\def\kaon    {{\ensuremath{\PK}}\xspace}
\def\Kp      {{\ensuremath{\kaon^+}}\xspace}
\def\Km      {{\ensuremath{\kaon^-}}\xspace}

\def\Dz      {{\ensuremath{\D^0}}\xspace}

\def\Dstarp  {{\ensuremath{\D^{*+}}}\xspace}

\def\Dsp     {{\ensuremath{\D^+_\squark}}\xspace}

\def\Dssp    {{\ensuremath{\D^{*+}_\squark}}\xspace}

\def\Bc      {{\ensuremath{\B_\cquark^+}}\xspace}

\def\jpsi     {{\ensuremath{{\PJ\mskip -3mu/\mskip -2mu\Ppsi}}}\xspace}

\renewcommand{\decay}[2]{\ensuremath{\mathinner{#1\!\to #2}}\xspace}

\def\to                 {\ensuremath{\rightarrow}\xspace}

\def\lplm     {\ensuremath{\ell^+ \ell^-}\xspace}

\def\BcJpsillDst {\decay{\Bc}{\jpsi(\to \ell^+ \ell^-) \Dstarp}}

\def\BcllDs {\decay{\Bc}{\Dsp\lplm}}

\def\BcllDst {\decay{\Bc}{\Dstarp\lplm}}

\def\BcJpsiDst {\decay{\Bc}{\jpsi \Dstarp}}
\def\BcJpsiDs {\decay{\Bc}{\jpsi \Dsp}}
\def\DKpi    {\decay{\Dz}{\Km \pip}}     
\def\DKpipipi    {\decay{\Dz}{\Km \pip \pip \pim}}  
\def\DKpipipi    {\decay{\Dz}{\Km \pip \pip \pim}}  
\def\DsKKpi     {\decay{\Dsp}{\Kp\Km \pip}}  
\def\DstDpi  {\decay{\Dstarp}{\Dz\pip}}  

\def\RDstDs {\ensuremath{\mathcal{R}_{\Dstarp / \Ds}}\xspace}

\def\BcJpsiPi {\decay{\Bc}{\jpsi \pip}}

\def\bcvegpy    {\mbox{\textsc{Bcvegpy}}\xspace}
\def\evtgen     {\mbox{\textsc{EvtGen}}\xspace}
\def\geant      {\mbox{\textsc{Geant4}}\xspace}
\def\photos     {\mbox{\textsc{Photos}}\xspace}
\def\pythia     {\mbox{\textsc{Pythia}}\xspace}

\usepackage{longtable} % only for template; not usually to be used in PAPERs

\usepackage[labelformat=simple]{subcaption}

\begin{document}

%%%%%%%%%%%%%%%%%%%%%%%%%
%%%%% Title     %%%%%%%%%
%%%%%%%%%%%%%%%%%%%%%%%%%
\renewcommand{\thefootnote}{\fnsymbol{footnote}}
\setcounter{footnote}{1}

% %%%%%%% CHOOSE TITLE PAGE--------
%\onecolumn
%\input{title-LHCb-INT}
%\input{title-LHCb-ANA}
%\input{title-LHCb-CONF}
%\input{title-LHCb-FIGURE}
% ===============================================================================
% Purpose: LHCb-PAPER journal paper title page template
% Author: 
% Created on: 2010-09-25
% ===============================================================================

%%%%%%%%%%%%%%%%%%%%%%%%%
%%%%%  TITLE PAGE  %%%%%%
%%%%%%%%%%%%%%%%%%%%%%%%%
\begin{titlepage}
\pagenumbering{roman}

% Header ---------------------------------------------------
\vspace*{-1.5cm}
\centerline{\large EUROPEAN ORGANIZATION FOR NUCLEAR RESEARCH (CERN)}
\vspace*{1.5cm}
\noindent
\begin{tabular*}{\linewidth}{lc@{\extracolsep{\fill}}r@{\extracolsep{0pt}}}
\ifthenelse{\boolean{pdflatex}}% Logo format choice
{\vspace*{-1.5cm}\mbox{\!\!\!\includegraphics[width=.14\textwidth]{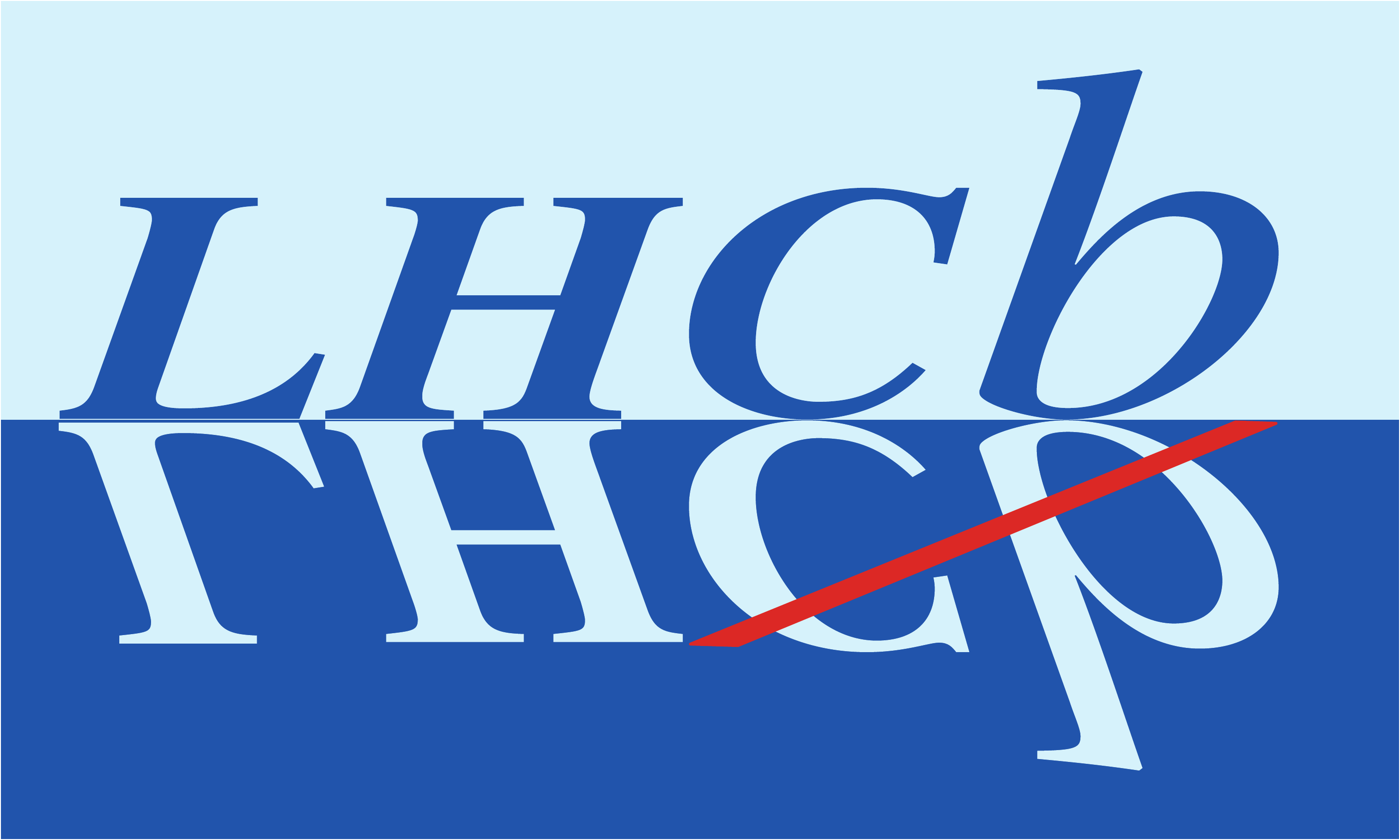}} & &}%
{\vspace*{-1.2cm}\mbox{\!\!\!\includegraphics[width=.12\textwidth]{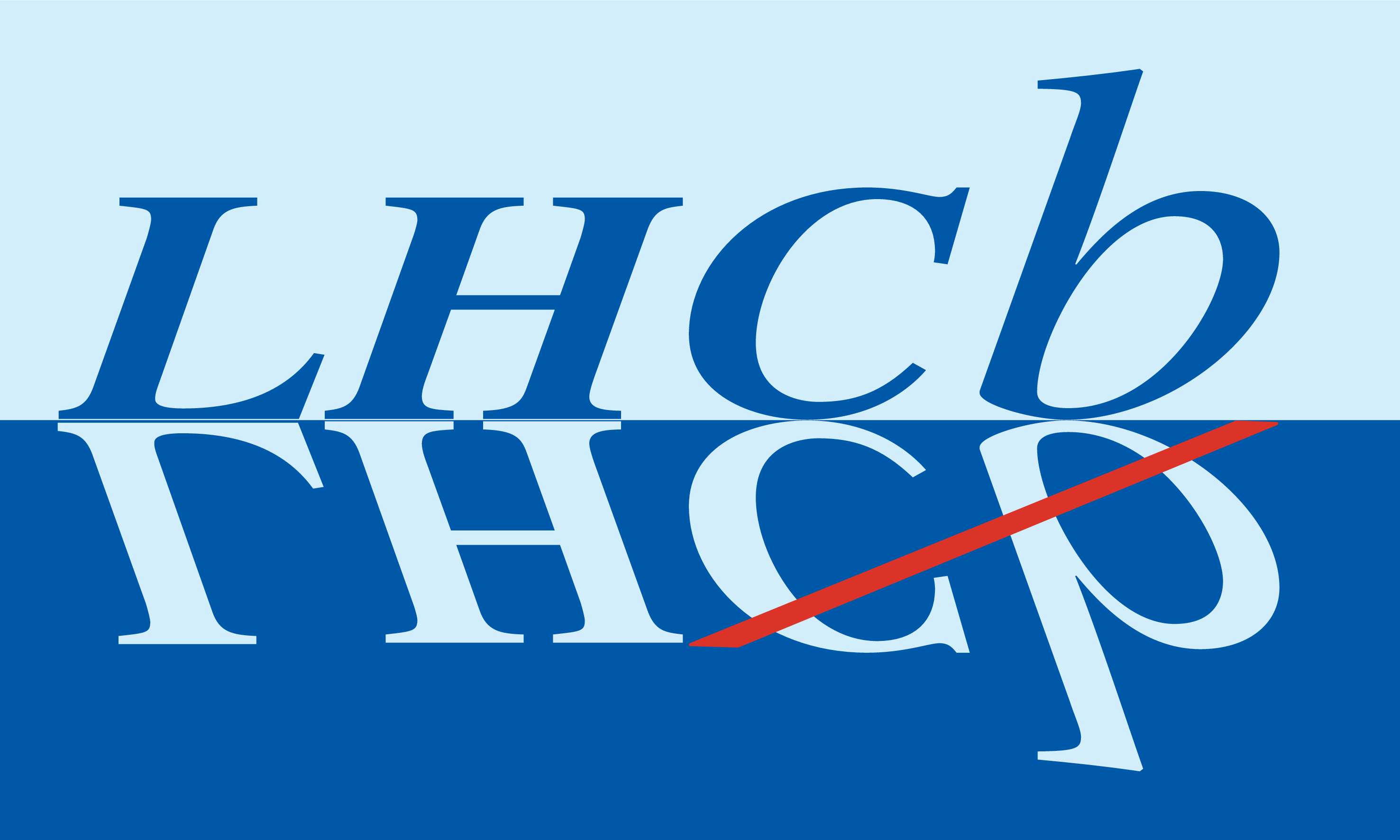}} & &}%
\\
 & & CERN-EP-2026-294 \\  % ID 
 & & LHCb-PAPER-2026-041 \\  % ID 
 & & November 18, 2026 \\ % Date - Can also hardwire e.g.: 23 March 2010
 & & \\
% not in paper \hline
\end{tabular*}

\vspace*{4.0cm}

% Title --------------------------------------------------
{\normalfont\bfseries\boldmath\huge
\begin{center}
% DO NOT EDIT HERE. Instead edit macro in main.tex to keep metadata correct
  \papertitle 
\end{center}
}

\vspace*{2.0cm}

% Authors -------------------------------------------------
\begin{center}
%In the footnote, replace 'paper' by 'Letter' in case of submission to PRL or PLB 
% Edit macro in main.tex to keep metadata correct
\paperauthors\footnote{Authors are listed at the end of this paper.}
\end{center}

\vspace{\fill}

% Abstract -----------------------------------------------
\begin{abstract}
  \noindent
  The first observation of the Cabibbo-suppressed decay \BcJpsiDst using proton-proton collision data collected by the \lhcb experiment between 2011 and 2018, corresponding to an integrated luminosity of 9\invfb, is presented. The observed signal has a significance of 5.4 standard deviations and the branching fraction relative to the normalisation mode \BcJpsiDs is measured to be 
  \begin{equation*}
    \frac{\mathcal{B}(\BcJpsiDst)}{\mathcal{B}(\BcJpsiDs)} = 0.067 \pm 0.016 \pm 0.003,
\end{equation*}
where the first uncertainty is statistical and the second is systematic. 
The result is used to test theoretical models of two-body hadronic decays of $\Bc$ mesons and agrees with several predictions in the literature.
\end{abstract}

\vspace*{2.0cm}
\begin{center}
  Submitted to JHEP 
\end{center}

\vspace{\fill}

{\footnotesize 
% Edit macro in main.tex to keep metadata correct
\centerline{\copyright~\papercopyright. \href{\paperlicenceurl}{\paperlicence}.}}
\vspace*{2mm}

\end{titlepage}

%%%%%%%%%%%%%%%%%%%%%%%%%%%%%%%%
%%%%%  EOD OF TITLE PAGE  %%%%%%
%%%%%%%%%%%%%%%%%%%%%%%%%%%%%%%%

%  empty page follows the title page ----
\newpage
\setcounter{page}{2}
\mbox{~}
%\newpage
%
%% Author List ----------------------------
%%  You need to get a new author list!
%\input{LHCb_authorlist.tex}
%

%\twocolumn
% %%%%%%%%%%%%% ---------

\renewcommand{\thefootnote}{\arabic{footnote}}
\setcounter{footnote}{0}

%%%%%%%%%%%%%%%%%%%%%%%%%%%%%%%%
%%%%%  Table of Content   %%%%%%
%%%%%%%%%%%%%%%%%%%%%%%%%%%%%%%%
%%%% Uncomment if desired
%\tableofcontents

\cleardoublepage

%%%%%%%%%%%%%%%%%%%%%%%%%
%%%%% Main text %%%%%%%%%
%%%%%%%%%%%%%%%%%%%%%%%%%

\pagestyle{plain} % restore page numbers for the main text
\setcounter{page}{1}
\pagenumbering{arabic}

%% Uncomment during review phase. 
%% Comment before a final submission.
% \linenumbers

\section{Introduction}
\label{sec:intro}
The \Bc meson is unique as it is the only flavoured pseudoscalar meson composed of two different heavy quarks and it remains the least explored ground state among the family of beauty mesons. While the production of \Bc mesons is suppressed with respect to the production of light beauty mesons $B_{(s)}$ by about two orders of magnitude, the LHCb experiment has collected a large dataset of \Bc meson decays, allowing for precise studies of its production, properties and decays~\cite{LHCb-PAPER-2012-054,LHCb-PAPER-2013-010,LHCb-PAPER-2013-044,LHCb-PAPER-2014-039,LHCb-PAPER-2014-050,LHCb-PAPER-2014-060,LHCb-PAPER-2019-007,LHCb-PAPER-2019-033,LHCb-PAPER-2020-003,LHCb-PAPER-2023-037,LHCb-PAPER-2023-046,LHCb-PAPER-2025-028,LHCb-PAPER-2024-035}.

Two-body decays of a \Bc meson to charmonium and a meson can be described using the quantum chromodynamics (QCD) factorisation approach, in which the decay amplitude separates into a decay constant and the \Bc to charmonium form factor. In contrast to the factorisation of $B_{(s)}$ decay amplitudes with light spectator quarks, which require nonperturbative \B-meson light-cone distribution amplitudes, \Bc decays with heavy spectator quarks require only the \Bc decay constant~\cite{Bordone:2022drp}. Therefore, precision measurements of nonleptonic two-body \Bc decays to charmonia allow for theoretically clean tests of the factorisation hypothesis and form-factor calculations from methods such as QCD sum rules~\cite{Kiselev:2002vz}, relativistic quark models~\cite{Anisimov:1998xv,Colangelo:1999zn,AbdEl-Hady:1999jux,Ivanov:2006ni} and perturbative QCD~\cite{Rui:2014tpa}. 

\begin{figure}[b]
\begin{subfigure}{0.3\textwidth}
\centering
    \includegraphics[width=\textwidth]{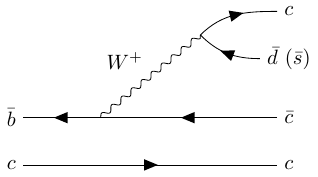}
    \caption{}
    \label{fig:bc_feynman_color_favoured}
\end{subfigure}
\hfill
\begin{subfigure}{0.3\textwidth}
\centering
    \includegraphics[width=\textwidth]{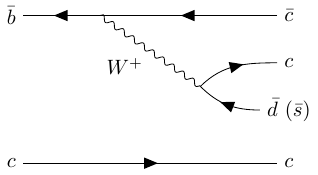}
    \caption{}
    \label{fig:bc_feynman_color_suppressed}
\end{subfigure}
\hfill
\begin{subfigure}{0.3\textwidth}
\centering
    \includegraphics[width=\textwidth]{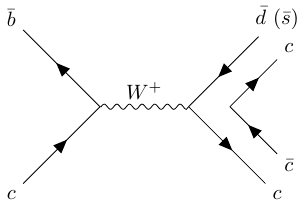}
    \caption{}
    \label{fig:bc_feynman_annihilation}
\end{subfigure}
\caption{Standard Model tree diagrams contributing to \BcJpsiDst (\BcJpsiDs) decays: (a) colour-allowed, (b) colour-suppressed and (c) annihilation diagrams.}
\label{fig:bc_feynman}
\end{figure}

In this context, the Cabibbo-suppressed decay \BcJpsiDst provides a sensitive probe for testing the theoretical description of exclusive \Bc decays, complementing the studies in \BcJpsiDs and \BcJpsiPi decays.\footnote{The inclusion of charge-conjugate processes is implied throughout.} As for the Cabibbo-favoured \BcJpsiDs decay, which has been studied previously by the \lhcb~\cite{LHCb-PAPER-2013-010,LHCb-PAPER-2022-048} and \atlas~\cite{ATLAS:2015jep,ATLAS:2022aiy} collaborations, the Cabibbo-suppressed decay \BcJpsiDst proceeds mainly through the tree diagrams illustrated in Figs.~\ref{fig:bc_feynman_color_favoured} and~\ref{fig:bc_feynman_color_suppressed}. Although the annihilation diagram shown in Fig.~\ref{fig:bc_feynman_annihilation} is enhanced for \Bc decays with respect to the analogous process in \Bp decays, the contribution for decays to charmonia and a charm meson is expected to be negligible due to a strong form-factor suppression~\cite{Colangelo:1999zn,Rui:2014tpa}.

\begin{figure}[t]
\centering
    \includegraphics[width=0.35\textwidth]{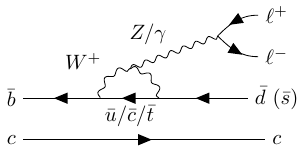}
\caption{Standard Model loop-level diagram contributing to a rare $b\to d\ell^+\ell^-$ ($b \to s\ell^+\ell^-$) transition in \BcllDst (\BcllDs) decays.}
\label{fig:bc_rare}
\end{figure}

Beyond serving as a probe of factorisation, the decay \BcJpsiDst is also valuable in light of future studies of flavour-changing neutral currents. Transitions of the type $b\to s\ell^+\ell^-$ or $b\to d\ell^+\ell^-$ are forbidden at tree level in the Standard Model and only proceed through higher-order loop-level diagrams as shown in Fig.~\ref{fig:bc_rare}. As a consequence, decays mediated by these transitions are rare, with typical branching fractions of $\mathcal{O}(10^{-6})$ or lower, and are a sensitive probe of New Physics, in which new particles and/or interactions would modify the decay amplitudes~\cite{Blake:2015tda,Altmannshofer:2021qrr}. 

At \lhcb, numerous analyses of $b \to s\ell^+\ell^-$ transitions have been performed~\cite{LHCb-PAPER-2024-046,LHCb-PAPER-2024-032,LHCb-PAPER-2024-011,LHCb-PAPER-2023-032,LHCb-PAPER-2022-045,LHCb-PAPER-2021-038} and extended to the study of $b \to d\ell^+\ell^-$ transitions~\cite{LHCb-PAPER-2012-020,LHCb-PAPER-2015-035,LHCb-PAPER-2018-004,LHCb-PAPER-2025-050}, which are an even more sensitive probe for New Physics due to the additional suppression of the involved Cabibbo--Kobayashi--Maskawa matrix element. As a result of the smaller production rate of \Bc mesons, observations of rare decays are so far restricted to light beauty hadrons, but complementary studies in \Bc decays will become accessible in the future. 
In this context, \BcllDst decays are particularly interesting as they offer a cleaner probe of \mbox{$b \to d \ell^+\ell^-$} transitions involving a vector resonance than the corresponding \mbox{${\decay{B^{+,0}}{\rho^{+,0}\ell^+\ell^-}}$} and \mbox{${\decay{\Bs}{K^{*0}\ell^+\ell^-}}$} decays, which suffer from large theoretical uncertainties due to broad light resonances or large experimental background from Cabibbo-favoured ${\decay{B^0}{K^{*0}\ell^+\ell^-}}$ decays, respectively.
To better constrain systematic uncertainties associated with the \bbbar production cross-section and experimental efficiencies, these analyses are typically performed relative to a control channel that decays to the same final state. For future studies of rare \BcllDst decays, the decay \BcJpsillDst provides optimal control of the experimental efficiencies, as it decays to the same final state, and its observation reported in this paper marks the first step towards the study of rare $b \to d \ell^+\ell^-$ decays of \Bc mesons.

This paper presents the first measurement of the \BcJpsiDst branching fraction relative to the more abundant \BcJpsiDs decay, defined as
\begin{equation}
    \RDstDs \equiv \frac{\mathcal{B}(\BcJpsiDst)}{\mathcal{B}(\BcJpsiDs)}.
\end{equation}
The two decays only differ by the presence of a vector or pseudoscalar charm meson with similar masses in the final state and by the involved Cabibbo--Kobayashi--Maskawa matrix elements. From the calculations in Refs.~\cite{Anisimov:1998xv,Colangelo:1999zn,AbdEl-Hady:1999jux,Kiselev:2002vz,Ivanov:2006ni,Rui:2014tpa}, which exploit different QCD descriptions of the \Bc meson, the ratio \RDstDs is predicted to be between 0.06 and 0.19. 

The analysis is based on a proton-proton ($pp$) data sample corresponding to an integrated luminosity of 9\invfb, collected by the LHCb experiment between 2011 and 2018. The dataset is divided into two data-taking periods: Run~1, corresponding to the 2011--2012 operation at centre-of-mass energies of 7 and 8\tev, and Run~2, corresponding to the 2015--2018 operation at a centre-of-mass energy of 13\tev. The \BcJpsiDst candidates are reconstructed from ${\decay{\jpsi}{\mumu}}$ and ${\decay{\Dstarp}{\Dz \pip}}$ decays with ${\decay{\Dz}{\Km\pip}}$ and ${\decay{\Dz}{\Km\pip\pip\pim}}$ decays. Candidates of the normalisation mode are reconstructed from ${\decay{\jpsi}{\mumu}}$ and ${\decay{\Dsp}{\Kp\Km\pip}}$. To avoid experimenter's bias, the invariant-mass region of the signal decay $\pm20\mevcc$ around the \Bc mass, corresponding to approximately three times the experimental resolution, was not examined until the full analysis procedure had been finalised.

\section{Detector and simulation}
\label{sec:DetSim}
The \lhcb detector~\cite{LHCb-DP-2008-001,LHCb-DP-2014-002} is a single-arm forward spectrometer covering the \mbox{pseudorapidity} range $2<\eta <5$, designed for the study of particles containing \bquark or \cquark quarks. The high-precision tracking system provides a measurement of the momentum, \ptot, of charged particles with a relative uncertainty that varies from 0.5\% at low momentum to 1.0\% at 200\gevc. The minimum distance of a track to a primary $pp$ collision vertex, the impact parameter, is measured with a resolution of $(15+29/\pt)\mum$, where \pt is the component of the momentum transverse to the beam, in\,\gevc. Different types of charged hadrons are distinguished using information from two ring-imaging Cherenkov detectors~\cite{LHCb-DP-2012-003}. Photons, electrons and hadrons are identified by a calorimeter system consisting of scintillating-pad and preshower detectors, an electromagnetic and a hadronic calorimeter. Muons are identified by a system composed of alternating layers of iron and multiwire proportional chambers~\cite{LHCb-DP-2012-002}.
The online event selection is performed by a trigger~\cite{LHCb-DP-2012-004,LHCb-DP-2019-001}, which consists of a hardware stage, based on information from the calorimeter and muon systems, followed by a software stage, which applies a full event reconstruction. At the hardware trigger stage, events are required to have a muon with high \pt or two oppositely charged muons with a high value of the product of their \pt. The software trigger identifies ${\decay{\jpsi}{\mumu}}$ candidates using good-quality tracks, identified as muons, which form a good-quality vertex and have a dimuon invariant mass within 120\mevcc of the known $\jpsi$ mass~\cite{PDG2026}.

Simulation is used to calculate reconstruction and selection efficiencies, and to determine shapes of invariant-mass distributions. In the simulation, the $pp$ collisions are generated using \pythia~\cite{Sjostrand:2007gs} with a specific LHCb configuration~\cite{LHCb-PROC-2010-056}. For the production of \Bc mesons, the \bcvegpy generator~\cite{Chang_2006} is used. 
Decays of unstable particles are described by \evtgen~\cite{Lange:2001uf}, incorporating final-state radiation generated using \photos~\cite{davidson2015photos}. The \BcJpsiDst signal decay is simulated using an amplitude model with equal magnitudes and no relative phases for the three helicity amplitudes, while the \BcJpsiDs normalisation decay is simulated using the angular distribution of a pseudoscalar decaying into a vector and a scalar. The \DKpipipi decays are generated according to the amplitude model from Ref.~\cite{LHCb-PAPER-2017-040}, and the \DsKKpi decay is generated with the Dalitz-plot model from Ref.~\cite{BaBar:2010wqe}. The interaction of the generated particles with the detector, along with its response, is implemented using the \geant toolkit~\cite{Allison:2006ve,*Agostinelli:2002hh} as described in Ref.~\cite{LHCb-PROC-2011-006}. 
The simulated samples are corrected to account for known differences between data and simulation in the $p$, \pt and $\eta$ distributions of the \Bc meson, as well as the track multiplicity of the $pp$ collision using a large sample of \BcJpsiPi decays. To account for imperfections in the simulation of charged-particle reconstruction, the track-reconstruction efficiency is corrected using data calibration samples of ${\decay{\jpsi}{\mumu}}$ decays~\cite{LHCb-DP-2013-002}. Imperfections in the particle identification (PID) response for charged hadron tracks are corrected using large data calibration samples of ${\decay{\Dstarp}{\Dz \pip}}$ with ${\decay{\Dz}{\Km \pip}}$ decays~\cite{LHCb-DP-2018-001}.

\section{\boldmath Candidate selection}
\label{sec:Sel}
The \BcJpsiDst candidates are formed by combining the \jpsi candidates that have been selected by the trigger requirements described in the previous section with a \Dstarp candidate. The latter is reconstructed through the decay \DstDpi with \DKpi and \DKpipipi, hereafter referred to as $2h$ and $4h$ decays, respectively. For the \BcJpsiDs normalisation mode, the \Dstarp candidate is replaced by a \DsKKpi candidate. 

All reconstructed tracks are required to be identified as kaons, pions or muons, to be of good quality and to be inconsistent with originating from any primary $pp$ collision vertex to ensure compatibility with the signal decays. Additional momentum requirements are imposed on all tracks to make sure they lie within the kinematic region for optimal PID~\cite{LHCb-DP-2018-001}. 

The \jpsi candidates are required to have an invariant mass within 30\mevcc of the known value~\cite{PDG2026}, whereas a $\pm 25\mevcc$ window for the \Dz and \Dsp, corresponding to approximately three standard deviations, is imposed. The difference between the invariant masses of the \Dstarp and \Dz candidates, $\Delta m$, is required to be in the range \mbox{143.5--147.5\mevcc}, exploiting the characteristic kinematics of the \DstDpi decay and covering approximately two times the experimental resolution. The residual nonresonant $\Dz\pip$ contribution is below the percent level and thus negligible. To suppress background from ${\decay{\Lc}{p \Km\pip}}$ decays in which the proton is misidentified as a kaon, \Dsp candidates are rejected if their invariant mass falls within 20\mevcc of the known mass of the \Lc baryon, when reconstructing the \Kp under the proton hypothesis. A kinematic fit~\cite{Hulsbergen:2005pu} is performed to each candidate, with vertex constraints applied to the \Bc, \jpsi, \Dstarp, \Dz and \Ds vertices, and with the \jpsi, \Dz and \Ds candidate masses constrained to their known values~\cite{PDG2026}. 

To suppress background stemming from random track combinations (combinatorial background), several boosted decision tree (BDT) classifiers~\cite{Breiman}, trained with the \texttt{XGBoost} algorithm~\cite{XGBoosting}, are employed. They are trained separately for the $2h$ and $4h$ signal channels, and the normalisation channel. 
To account for differences in the decay distributions between Run~1 and Run~2, the training and optimisation of the BDTs are performed separately for the two run periods. 
Data from the upper mass sideband in the range 6400--7000\mevcc serve as a combinatorial background proxy. Due to the lower level of combinatorial background in the $2h$ signal channel, the sideband data are enlarged by relaxing the requirement on $\Delta m$ to $141.5 < \Delta m < 149.5\mevcc$. The simulated samples of the three decay modes are used as the respective signal proxy. A $k$-fold cross-validation approach with $k = 5$ is used~\cite{kFold}, in which each of the five classifiers is trained leaving out a different 20\% of the training sample. This ensures no candidate is evaluated by a classifier trained on it.
The classifiers for the \BcJpsiDst and \BcJpsiDs channels are trained on similar sets of decay variables, 
including information on the decay topology, kinematics and vertex-fit quality of the \Bc candidate and the quality of the aforementioned kinematic fit with additional vertex and mass constraints.

For the \BcJpsiDst decays, the optimal requirements for the BDT are determined by maximising the figure of merit defined in Ref.~\cite{Punzi:2003bu} for a signal significance of \mbox{five standard deviations}, $\varepsilon_\text{sig}/(5/2+\sqrt{N_\text{bkg}})$, expected for the signal region \mbox{$6255 < m(\jpsi\Dz\pi^+) < 6295\mevcc$}. Here, $\varepsilon_\text{sig}$ is the signal efficiency of the corresponding BDT requirement, and $N_\text{bkg}$ is the expected number of background candidates in the signal window extrapolated from a fit to the data sideband for a given BDT requirement. With the optimal requirement, the BDT reduces the background level by 97\% and 99.9\% for the $2h$ and $4h$ signal decays, respectively, while retaining 79\% and 75\% of the signal. For the far more abundant normalisation mode \BcJpsiDs, the BDT requirement is optimised by maximising the value of $N_\text{sig}^\text{norm}/\sqrt{N_\text{sig}^\text{norm}+N_\text{bkg}^\text{norm}}$ in the signal window, where $N_\text{sig}^\text{norm}$ is the expected number of \BcJpsiDs decays. 
The optimal BDT requirement reduces the background level by 99.8\% and retains 73\% of the \BcJpsiDs candidates. After the whole selection, about 2\% of the events contain multiple candidates in the signal and normalisation channels, in which case one candidate is chosen at random. 

The efficiency of the candidate selection is determined using simulated \BcJpsiDst and \BcJpsiDs decays. The trigger efficiency is corrected using a large sample of ${\decay{\Bp}{\jpsi\Kp}}$ decays and the method described in Ref.~\cite{LHCb-DP-2012-004}. 
The ratios of the total reconstruction and selection efficiencies between the signal and normalisation decays are $\varepsilon_{2h}/\varepsilon_{\Ds}=0.772\pm 0.005$ and $\varepsilon_{4h}/\varepsilon_{\Ds}=0.2796\pm 0.0035$, where the uncertainties are from the finite size of the simulated samples only. 

\section{\boldmath Determination of the \BcJpsiDs yield \label{sec:DsFit}}
The yield of the \BcJpsiDs normalisation mode is determined through an extended unbinned maximum-likelihood fit to the distribution of the reconstructed invariant mass, $m(\jpsi \Dsp)$. The candidate samples from different years of data taking are combined in the fit.

The invariant-mass distribution of signal candidates is modelled by the sum of a double-sided Crystal Ball function (DSCB)~\cite{Skwarnicki:1986xj}, comprising a Gaussian core and power-law tails on both sides of the peak, and a Gaussian component. The mean of the two components is the same. The parameters for the shape of the \BcJpsiDs candidates are determined from simulation and fixed in the fit, except for a width scale factor and mass shift, which are free parameters to account for differences between data and simulation. 

\begin{figure}[tb]
    \centering
    \includegraphics[width=0.45\linewidth]{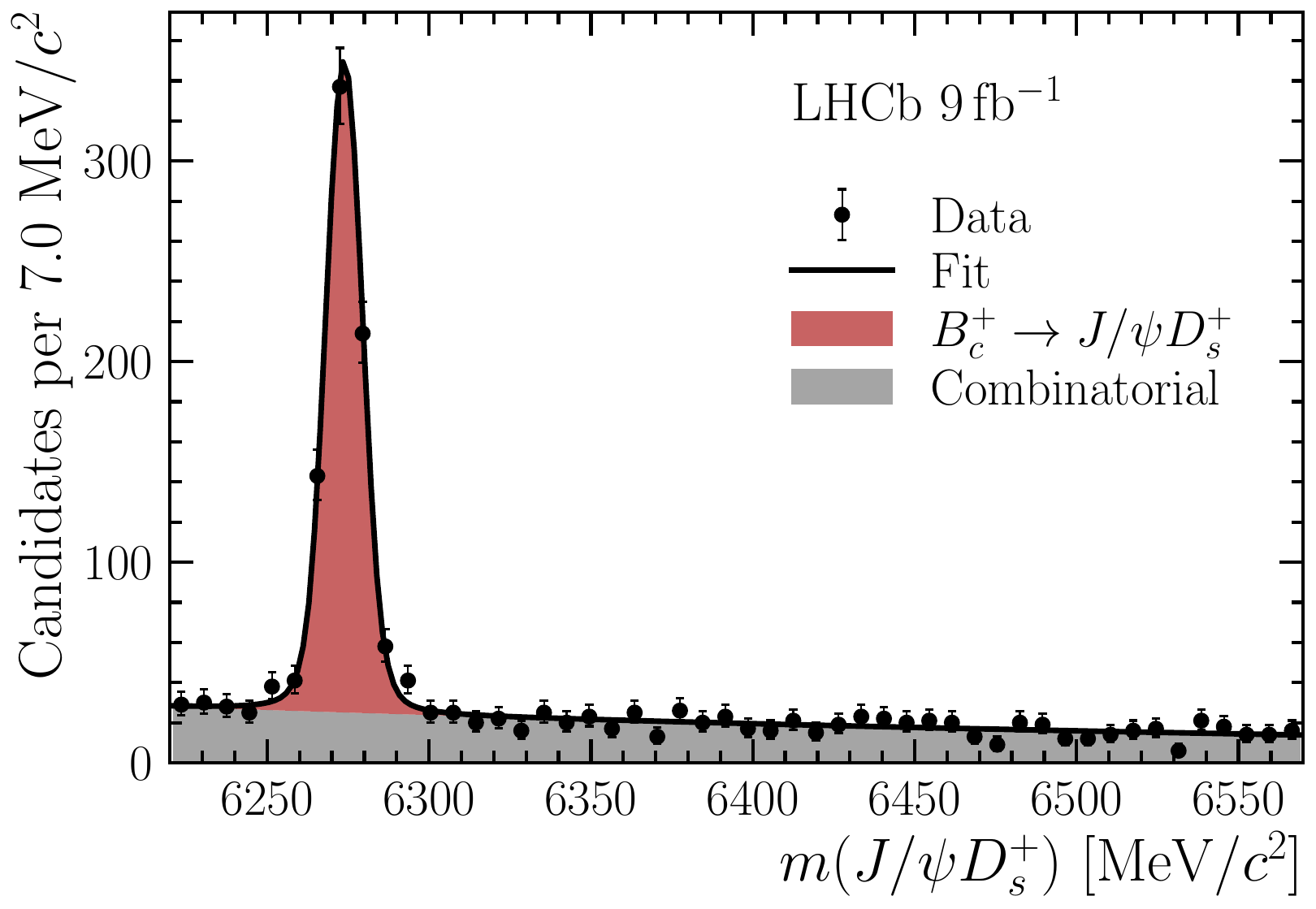}
    \caption{Invariant-mass distribution of \BcJpsiDs candidates with \DsKKpi in the combined Run~1 and Run~2 data and with the fit result overlaid.}
    \label{fig:fit_Ds}
\end{figure}

The combinatorial background is described by an exponential function whose slope is a free parameter in the fit. The fit to $m(\jpsi \Dsp)$ is shown in Fig.~\ref{fig:fit_Ds}. The signal yield in the normalisation channel is found to be $N_{\Dsp}=696\pm30$. 

\section{\boldmath Invariant-mass fits of \BcJpsiDst candidates}
\begin{figure}[tb]
    \centering
    \includegraphics[width=0.45\textwidth]{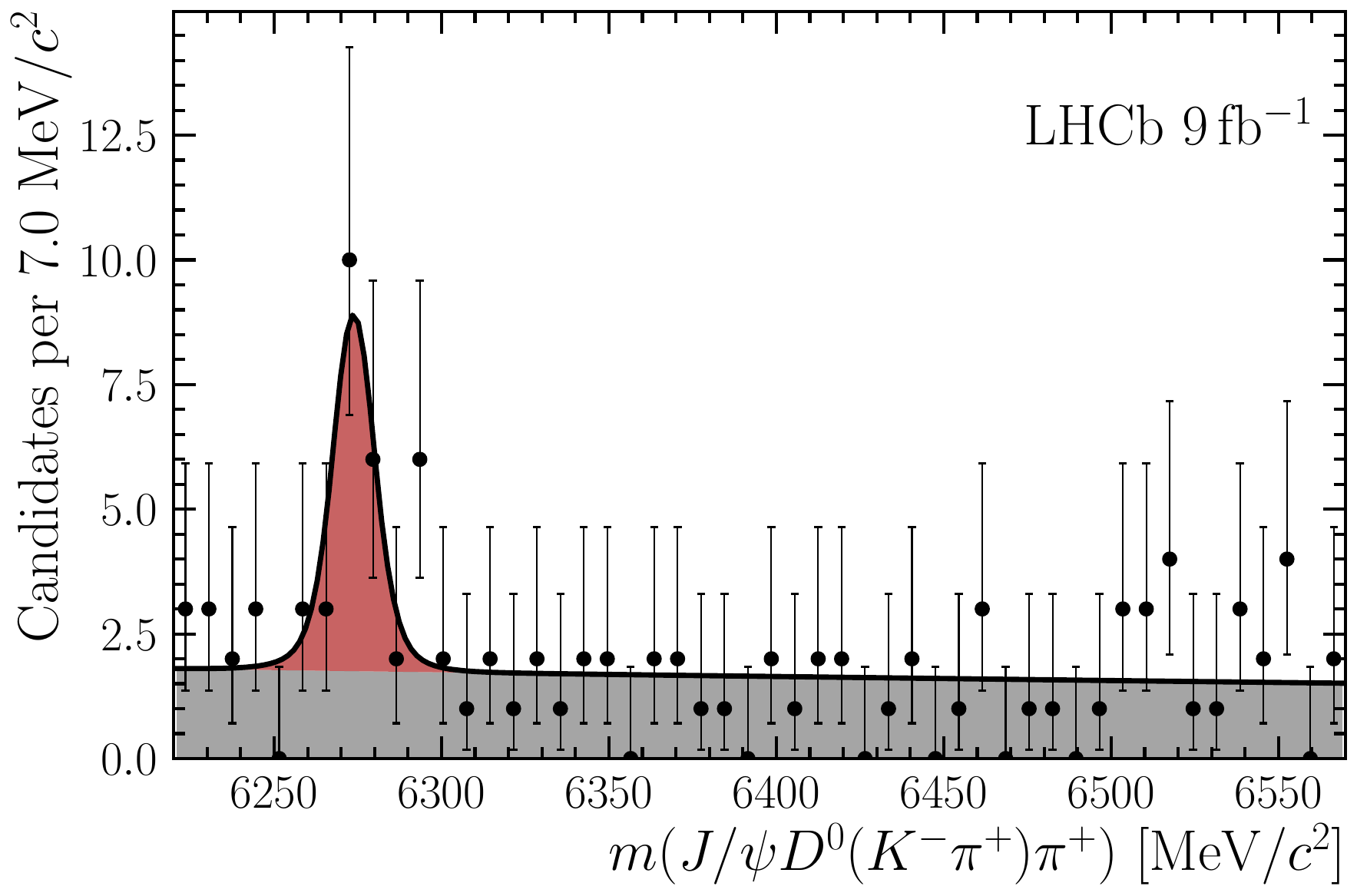}
    \hfill
    \includegraphics[width=0.45\textwidth]{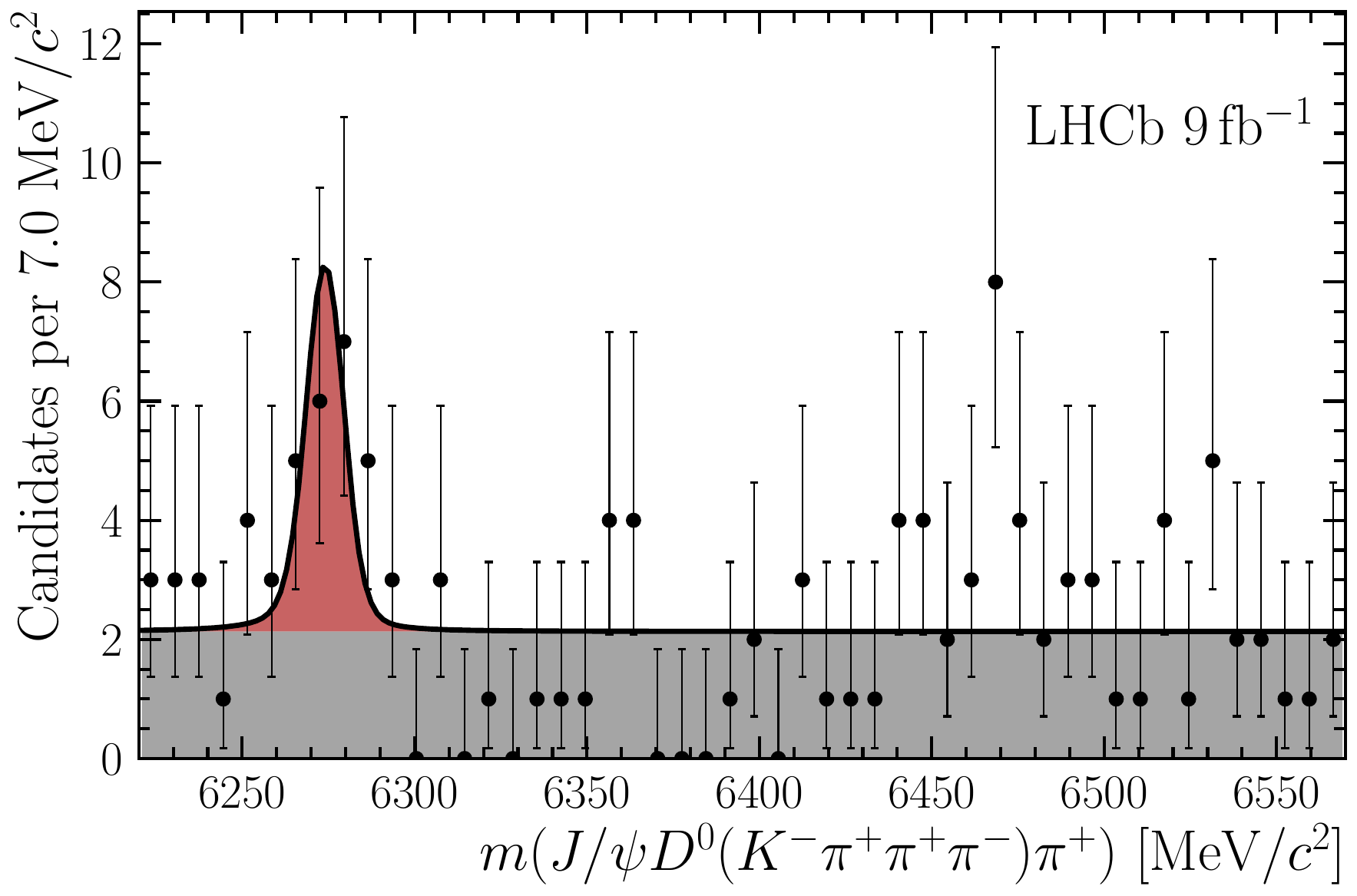}\\
    \includegraphics[width=0.56\textwidth]{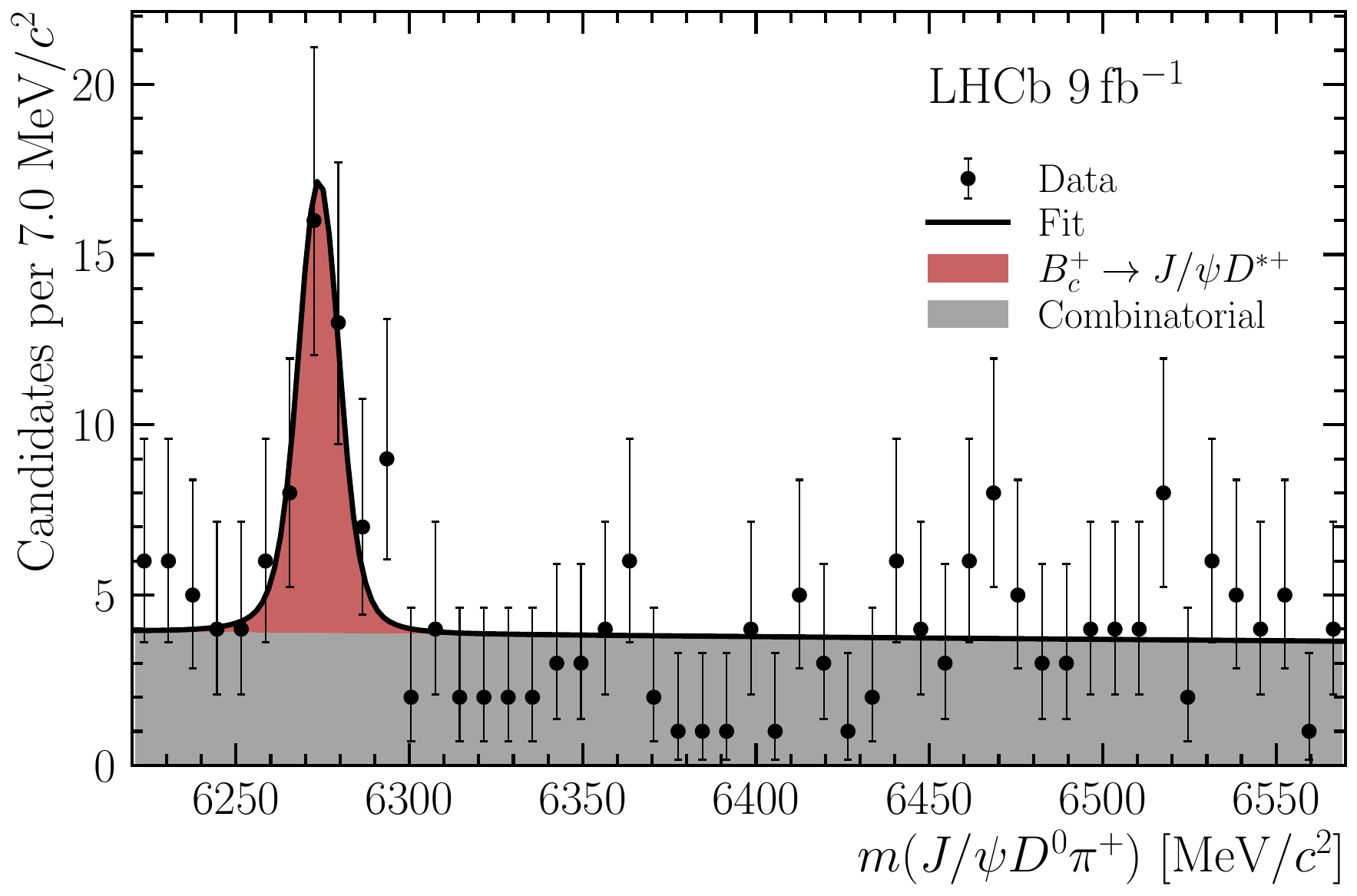}
    \caption{Invariant-mass distributions of \BcJpsiDst candidates with (top left) \DKpi, (top right) \DKpipipi decays and (bottom) both combined. The result of the simultaneous fit is overlaid.}
    \label{fig:DstarFit}
\end{figure}

An extended unbinned maximum-likelihood fit to the distribution of the reconstructed invariant mass, $m(\jpsi \Dz \pip)$, of \BcJpsiDst candidates is performed simultaneously for the $2h$ and $4h$ decay modes to determine the ratio of branching fractions \RDstDs. For each \Dz decay mode, the signal shape is described by the sum of a DSCB and a Gaussian function with a shared mean. The model parameters are fixed from simulation and the widths and mean positions are corrected by the scale and shift factors obtained from the fit of the normalisation mode and are Gaussian-constrained to account for the respective uncertainties. 
The combinatorial background is described by an exponential function and the slope is fitted separately from the $2h$ and $4h$ data. 
For each \Dz final state $f=2h,4h$, the signal yield $N_f$ is expressed in terms of the ratio of branching fractions \RDstDs as
\begin{equation}
\label{eq:signal}
    N_f = \RDstDs \times N_{\Ds} \times \frac{\varepsilon_{f}}{\varepsilon_{\Ds}}\times \frac{\mathcal{B}( \DstDpi)\times \mathcal{B}(\Dz \to f)}{\mathcal{B}(\DsKKpi)}.
\end{equation}
Here, $N_{D_s^+}$ is the yield of the normalisation mode given in the previous section, $\varepsilon_f/\varepsilon_{D_s^+}$ is the ratio of efficiencies of the signal and normalisation decays reported in Sec.~\ref{sec:Sel} and the external inputs on the charm-meson branching fractions are taken from Ref.~\cite{PDG2026}. All parameters are fixed to their central values and the respective uncertainties are discussed as sources of systematic uncertainty in the following section. The observed signal yields are $N_{2h}=18\pm4$ and $N_{4h}=13\pm3$, respectively. This fitting procedure is validated through the fit of pseudoexperiments that include both signal and background components with yields obtained from the data fit. A small negative bias, about an order of magnitude smaller than the statistical uncertainty, is observed due to the limited sample size. This bias is assigned as a systematic uncertainty.
The fit projections for the two different final states, along with the combination of all final states, are shown in Fig.~\ref{fig:DstarFit}. The ratio of branching fractions is measured to be $\RDstDs = 0.067 \pm 0.016$, where the uncertainty is statistical only and dominated by the uncertainties on the \BcJpsiDst signal yields. 

\section{Systematic uncertainties \label{sec:Syst}} 
Many of the systematic uncertainties cancel in the ratio of efficiencies used in Eq.~\ref{eq:signal}. Nevertheless, several contributions remain and are summarised in Table~\ref{tab:systematics}. 

The dominant systematic uncertainty arises from the calibration of the reconstruction and selection efficiency in simulation. The finite size of the data control samples used to calibrate the track-reconstruction and PID efficiencies introduces a systematic uncertainty, which, due to the different final states of the signal and normalisation decays, does not fully cancel. It is propagated to the ratio of branching fractions using a bootstrapping method~\cite{efron:1979}. The same method is also used to propagate the uncertainties linked to the calibration of the trigger response. The modelling of nuclear interactions in the simulation introduces an additional uncertainty on the efficiency calibration due to the uncertainty on the material distribution, which is accounted for using the procedure outlined in Ref.~\cite{LHCb-DP-2013-002}. 

A systematic uncertainty covering the size of the small bias observed during the validation of the fit is also added. 

A further systematic uncertainty arises from the angular model used in the simulation of \BcJpsiDst decays, based on which the signal selection efficiency is evaluated. The effect of a different polarisation of the \Bc decay products is studied by recalculating the efficiency with a purely longitudinally and a purely transversely polarised simulation sample, separately for the $2h$ and $4h$ decays. The larger of the two deviations from the nominal efficiency is assigned as a systematic uncertainty on the respective selection efficiency. The impact on the value of \RDstDs is examined in pseudoexperiments, in which the $2h$ and $4h$ signal efficiencies are varied jointly within the respective uncertainties. 
 
The uncertainties on the external branching fractions~\cite{PDG2026} are propagated to \RDstDs. 

Small differences in the number of multiple candidates in data and simulation introduce an uncertainty on the efficiency of the candidate removal. It is quantified by fitting the \BcJpsiDs data without removing multiple candidates and propagating the uncertainty on the ratio of branching fractions. For a conservative estimate, no cancellation between the signal and normalisation is assumed. As the rate of multiple candidates in the simulation of $4h$ signal decays is a factor of 1.5 larger than for the normalisation decay, the assigned systematic uncertainty is scaled by the same factor.

Another source of systematic uncertainty arises from the choice of models used to describe the signal and background components in the fit of \BcJpsiDs and \BcJpsiDst candidates. Its impact on the ratio of branching fractions is evaluated using pseudoexperiments generated from an alternative model, in which the signal is parametrised by a generalised Crystal Ball function with independent widths on the left- and right-hand sides, while the background is described by a second-order polynomial. These pseudodata, constructed to have sample sizes comparable to those of the data, are subsequently fitted using the nominal model.

Lastly, the finite size of the simulation samples generated for the decays \BcJpsiDs and \BcJpsiDst introduces an uncertainty on the derived efficiency ratios, which is propagated to the final ratio of branching fractions. 

\begin{table}[tb]
    \centering
    \caption{Systematic uncertainties in the determination of the \RDstDs ratio.}
    \label{tab:systematics}
    \begin{tabular}{lc}
        \toprule
        Systematic uncertainty & $\Delta\RDstDs [10^{-3}]$ \\
        \midrule
        Efficiency calibration & 2.0\\
        Fit bias & 1.5\\
        \BcJpsiDst decay model & 1.4\\
        External uncertainties & 1.3\\
        Multiple candidates & 0.8\\
        Fit model & 0.7\\
        Finite simulation samples & 0.5\\
        \midrule
        Total systematic uncertainty & 3.3\\
        Statistical uncertainty & 16.2\phantom{0}\\
        \bottomrule
    \end{tabular}
   
\end{table}

\section{Signal significance}
\label{sec:Res}
To estimate the signal significance, a total of 10\,000 pseudoexperiments are generated under the background-only hypothesis and subsequently fitted with a background-only and signal-plus-background model. From the likelihood ratio $q=-2\ln (\mathcal{L}_{B} / \mathcal{L}_{S+B})$, the cumulative probability distribution $\mathcal{P}(q \geq q_{\rm obs})$ is obtained as a function of the observed likelihood ratio, $q_{\rm obs}$. It is modelled with a $\chi^2$ distribution, where the effective number of degrees of freedom is obtained from a fit. Subsequently, the tail of this distribution is extrapolated to the observed likelihood ratio from the fit to the data to estimate a $p$-value, which corresponds to a significance of 5.4 standard deviations including systematic uncertainties. Consequently, this marks the first observation of the decay \BcJpsiDst. 

\section{Summary}
\label{sec:Summary}
The decay \BcJpsiDst is observed for the first time with a significance of 5.4 standard deviations by employing $pp$ collision data collected with the \lhcb experiment corresponding to an integrated luminosity of 9\invfb. The measured branching fraction relative to the normalisation mode \BcJpsiDs is 
\begin{equation*}
\label{eq:signal_theory}
    \RDstDs = 0.067 \pm 0.016 \pm 0.003,
\end{equation*}
where the first uncertainty is statistical and the second one is systematic. The value is in good agreement with the theoretical predictions of 0.056 and 0.066 from Ref.~\cite{Colangelo:1999zn} using the values of the Wilson coefficients from Refs.~\cite{Colangelo:1999zn} and~\cite{Ivanov:2006ni}, respectively. This is consistent with the agreement found in the Cabibbo-favoured decays \BcJpsiDs and ${\decay{\Bc}{\jpsi\Dssp}}$ by the \lhcb~\cite{LHCb-PAPER-2013-010,LHCb-PAPER-2022-048} and \atlas~\cite{ATLAS:2015jep,ATLAS:2022aiy} collaborations. 
The models based on the light-front constituent quark model from Ref.~\cite{Anisimov:1998xv} and perturbative QCD from Ref.~\cite{Rui:2014tpa}, which both predict \RDstDs to be 0.083, also agree well with the measured value within the experimental uncertainties.
The relativistic quark models in Refs.~\cite{AbdEl-Hady:1999jux,Ivanov:2006ni}, with which \RDstDs is predicted to be 0.19 and 0.13, respectively, and the model based on QCD sum rules in Ref.~\cite{Kiselev:2002vz}, predicting a value of 0.16, are disfavoured.

Since 2024, the upgraded \lhcb experiment has already recorded a dataset containing approximately three times as many \bbbar
events as those analysed in this publication. This substantially larger sample will enable more stringent tests of QCD in the doubly heavy-quark system of the \Bc meson. 
With the accumulation of further data, the first observation of the $\bquark \to \squark \ell^+ \ell^-$ transition in \Bc decays may become possible. Furthermore, the first observation of the \BcJpsiDst control mode, reported in this publication, is an important step towards the first search for the even rarer \BcllDst decay.
\section*{Acknowledgements}
%
% These Acknowledgements valid from 16/06/2026
%
\noindent We express our gratitude to our colleagues in the CERN
accelerator departments for the excellent performance of the LHC. We
thank the technical and administrative staff at the LHCb
institutes.
We acknowledge support from CERN and from the national agencies:
ARC (Australia);
CAPES, CNPq, FAPERJ and FINEP (Brazil); 
MOST and NSFC (China); 
CNRS/IN2P3 and CEA (France);  % added CEA 26/02/2026
BMFTR, DFG and MPG (Germany);
NKFIH (Hungary);              % added 16/06/2026
INFN (Italy); 
NWO (Netherlands); 
MNiSW and NCN (Poland); 
MEC/IFA (Romania); 
%MSHE (Russia); 
MICIU and AEI (Spain);
SNSF and SER (Switzerland); 
NASU (Ukraine); 
STFC (United Kingdom); 
DOE NP and NSF (USA).
%%%%%%%%%%%%%%%%%%%%%%%%%%%%%%%%%%%%%%%%%%%%%
We acknowledge the computing resources that are provided by ARDC (Australia), 
CBPF (Brazil),
CERN, 
IHEP and LZU (China),
IN2P3 (France), 
KIT and DESY (Germany), 
INFN (Italy), 
SURF (Netherlands),
Polish WLCG (Poland),
IFIN-HH (Romania), % http://dx.doi.org/10.13039/100019931,"Institutul National de Cercetare-Dezvoltare pentru Fizica si Inginerie Nucleara 'Horia Hulubei'"
%RRCKI and Yandex LLC (Russia), 
PIC (Spain), CSCS (Switzerland), 
GridPP (United Kingdom),
and NSF (USA).  % added Feb2026
%%%%%%%%%%%%%%%%%%%%%%%%%%%%%%%%%%%%%%%%%% 
We are indebted to the communities behind the multiple open-source
software packages on which we depend.
%%%%%%%%%%%%%%%%%%%%%%%%%%%%%%%%%%%%%%%%%%
Individual groups or members have received support from
% ARC and ARDC (Australia); % moved to national 16/01/2025
RTP (Australia), % added 06/03/2026
FWO Odysseus grant G0ASD25N (Belgium), % added 20/4/2026
Key Research Program of Frontier Sciences of CAS, CAS PIFI, CAS CCEPP (China); 
%Fundamental Research Funds for the Central Universities,  and Sci.\
%\& Tech.\ Program of Guangzhou (China); Removed 24/11/25
Minciencias (Colombia);
EPLANET, Marie Sk\l{}odowska-Curie Actions, ERC and NextGenerationEU (European Union);
A*MIDEX, ANR, IPhU and Labex P2IO, and R\'{e}gion Auvergne-Rh\^{o}ne-Alpes (France);
%RFBR, RSF and Yandex LLC (Russia);
Alexander-von-Humboldt Foundation (Germany);
ICSC (Italy); 
%GVA, XuntaGal, GENCAT, Inditex, InTalent and Prog.~Atracci\'on Talento, CM (Spain);
Severo Ochoa and Mar\'ia de Maeztu Units of Excellence, GVA, XuntaGal, GENCAT, InTalent-Inditex and Prog.~Atracci\'on Talento CM (Spain);
%XuntaGal --> Xunta de Galicia 
% SRC (Sweden);  % removed 27/02/2026 - end of grant
the Leverhulme Trust, the Royal Society and UKRI (United Kingdom).

%\input{LHCbInternal/supplementary_template}

%\input{appendix}

% This should be taken out in the final paper
%\input{supplementary}

\addcontentsline{toc}{section}{References}
%\setboolean{inbibliography}{true}
\bibliographystyle{LHCb/LHCb}
\bibliography{main,LHCb/standard,LHCb/LHCb-PAPER,LHCb/LHCb-CONF,LHCb/LHCb-DP,LHCb/LHCb-TDR,LHCb/LHCb-PUB}
 
\newpage
% LHCb collaboration author list
% Data extracted on November 13th, 2026 at 3:52pm for paper reference LHCb-PAPER-2026-041
\centerline
{\large\bf LHCb collaboration}
\begin
{flushleft}
\small
R.~Aaij$^{41}$\lhcborcid{0000-0003-0533-1952},
M.~Abdelfatah$^{73}$,
A.S.W.~Abdelmotteleb$^{61}$\lhcborcid{0000-0001-7905-0542},
C.~Abellan~Beteta$^{55}$\lhcborcid{0009-0009-0869-6798},
F.~Abudin\'en$^{63}$\lhcborcid{0000-0002-6737-3528},
T.~Ackernley$^{65}$\lhcborcid{0000-0002-5951-3498},
A.A.~Adefisoye$^{73}$\lhcborcid{0000-0003-2448-1550},
B.~Adeva$^{51}$\lhcborcid{0000-0001-9756-3712},
M.~Adinolfi$^{59}$\lhcborcid{0000-0002-1326-1264},
P.~Adlarson$^{90,46}$\lhcborcid{0000-0001-6280-3851}
% ...

{\footnotesize \it

% ...
$^{41}$Nikhef National Institute for Subatomic Physics, Amsterdam, Netherlands\\
% ...
$^{46}$National Center for Nuclear Research (NCBJ), Warsaw, Poland\\
% ...
$^{51}$Instituto Galego de F{\'\i}sica de Altas Enerx{\'\i}as (IGFAE), Universidade de Santiago de Compostela, Santiago de Compostela, Spain\\
% ...
$^{55}$Physik-Institut, Universit{\"a}t Z{\"u}rich, Z{\"u}rich, Switzerland\\
% ...
$^{59}$H.H. Wills Physics Laboratory, University of Bristol, Bristol, United Kingdom\\
% ...
$^{61}$Department of Physics, University of Warwick, Coventry, United Kingdom\\
% ...
$^{63}$School of Physics and Astronomy, University of Edinburgh, Edinburgh, United Kingdom\\
$^{64}$School of Physics and Astronomy, University of Glasgow, Glasgow, United Kingdom\\
$^{65}$Oliver Lodge Laboratory, University of Liverpool, Liverpool, United Kingdom\\
% ...
$^{73}$Syracuse University, Syracuse, NY, United States\\
% ...
$^{90}$Department of Physics and Astronomy, Uppsala University, Uppsala, Sweden, associated to $^{64}$\\
% ...
}
\end{flushleft}
\end{document}